\documentstyle[12pt]{article}\textheight 220mm\textwidth 160mm
\begin{document}

\def\Journal#1#2#3#4{{#1} {\bf #2}, #3 (#4)}

\def\NCA{\em Nuovo Cimento}
\def\NIM{\em Nucl. Instrum. Methods}
\def\NIMA{{\em Nucl. Instrum. Methods} A}
\def\NPB{{\em Nucl. Phys.} B}
\def\PLB{{\em Phys. Lett.}  B}
\def\PRL{\em Phys. Rev. Lett.}
\def\PRD{{\em Phys. Rev.} D}
\def\ZPC{{\em Z. Phys.} C}

\def\st{\scriptstyle}
\def\sst{\scriptscriptstyle}
\def\mco{\multicolumn}
\def\epp{\epsilon^{\prime}}
\def\vep{\varepsilon}
\def\ra{\rightarrow}
\def\ppg{\pi^+\pi^-\gamma}
\def\vp{{\bf p}}
\def\ko{K^0}
\def\kb{\bar{K^0}}
\def\al{\alpha}
\def\ab{\bar{\alpha}}
\def\be{\begin{equation}}
\def\ee{\end{equation}}
\def\bea{\begin{eqnarray}}
\def\eea{\end{eqnarray}}
\newcommand{\bi}{\bibitem}

\catcode`\@=11
\def\lsim{\mathrel{\mathpalette\@versim<}}
\def\gsim{\mathrel{\mathpalette\@versim>}}
\def\@versim#1#2{\vcenter{\offinterlineskip
\ialign{$\m@th#1\hfil##\hfil$\crcr#2\crcr\sim\crcr } }}
\catcode`\@=12
\renewcommand{\baselinestretch}{1.5}
\pagestyle{empty}
\newcommand{\nn}{\nonumber}

\noindent
\hspace*{10.7cm} \vspace{-2mm} KANAZAWA-97-14

\begin{center}UNIFICATION BEYOND 
GUTS--TOP MASS PREDICTIONS $^{\dag,*}$ \\

\vspace{2cm}

Jisuke KUBO \\

Physics Dept., Kanazawa University, Kanazawa, 920-11 Japan \\

\vspace{0.2cm}
 Myriam MONDRAGON \\

Depto. de F\'{\i}sica Te\'orica, Instituto de F\'{\i}sica,
  UNAM,  M\'exico 01000 D.F.\\

\vspace{0.2cm} 
George ZOUPANOS$ ^{**}$\\

Instit\" ut f\" ur Physik, Humboldt-Universit\" at zu Berlin,
D-10115 Berlin, Germany

\end{center}

\vspace{1.5cm}
\begin{center}
ABSTRACT
\end{center}

\vspace{0.5cm}
Unification of Gauge and Yukawa sectors of GUTs is obtained
by searching for renormalization group invariant relations
among the couplings of the both sectors which hold beyond the
unification scale.
This procedure singled out two supersymmetric GUTs, the finite and the
minimal $SU(5)$ models which, among others, were predicting successfully
 the top quark mass. The same procedure is currently 
extended in the soft supersymmetry-breaking sector of the theories,
 providing us with interesting, though preliminary, predictions on the
Higgs and superpartner masses.

\vspace*{1.5cm}
\footnoterule
\vspace*{2mm}
\noindent

$ ^{\dag}$ Presented by G. Zoupanos at
{\em  32nd Recontres de Moriond 
 on Electroweak Interactions and Unified Theories}, 
Les Arcs, France, March 15-22, 1997, 
to appear in the proceedings.\\
$^{*}$ Partially supported  by the E.C. projects, 
FMBI-CT96-1212 and ERBFMRXCT960090,
the Greek projects, PENED95/1170; 1981,
the UNAM Papiit project IN110296, 
and the Grants-in-Aid
for Scientific Research  from the Ministry of
Education, Science 
and Culture in Japan, No. 40211213.\\
$ ^{**}$ On leave from:
Physics Dept., Nat. Technical University, GR-157 80 Zografou,
  Athens, Greece

\newpage
\pagestyle{plain}

\section{Introduction}
Unification of all interactions has always been a very attractive 
theoretical dream, and 
has proved  in certain cases to be a powerful tool in our endeavor 
to reduce the plethora of free parameters of the Standard Model (SM). For
instance,  unification of the SM forces based on the  $SU(5)$ GUT was predicting
one of the  gauge couplings as well as the mass of the bottom quark.
Now it seems that LEP data is suggesting that the symmetry of the
unified theory should be further enlarged and become  $N=1$ globally
supersymmetric.

Relations among gauge and Yukawa couplings,
which are missing in ordinary GUTs,
could be a consequence of a further unification provided by a more 
fundamental theory
at the Planck scale. Moreover,  it might be possible that
some of these relations are renormalization group invariant (RGI)
below the  Planck scale so that they  
are exactly preserved down to the GUT scale.
\footnote{An indication has been recently found
in the soft supersymmetry-breaking sector as we will mention later.}
In our recent studies \cite{kubo2},
we have been searching for
RGI relations among gauge and Yukawa couplings in various GUTs.
Two of our models are of particular interest.
One based on the minimal supersymmetric $SU(5)$ is very attractive 
due to its simplicity.
The other, which is a finite $SU(5)$ is exciting, since it is
a realization of another old theoretical dream, namely that a truly
unified theory should not have infinities.
We should stress that the latter is a common view shared by many theoreticians
including those working on strings, 
non-commutative geometry and
quantum groups. Since it is widely expected  that it is necessary to
include gravity in the unified picture in order to achieve
finiteness, we hope it is interesting for many to know that  
models which are finite to all orders in perturbation theory can be constructed,
without introducing gravitational interactions, 
 by searching for RGI
relations beyond unification scale among gauge and Yukawa couplings. 

Moreover both models among other successful postdictions
provided us with predictions of
the top quark mass, which have passed successfully so 
far all tests of progressively more accurate measurements.

\section{ Gauge-Yukawa Unification}

A RGI relation among couplings,
$\Phi (g_1,\cdots,g_N) ~=~0$,
has to satisfy the partial differential equation (PDE)
$\mu\,d \Phi /d \mu ~=~
\sum_{i=1}^{N}
\,\beta_{i}\,\partial \Phi /\partial g_{i}~=~0$, 
where $\beta_i$ is the $\beta$-function of $g_i$.
There exist ($N-1$) independent  $\Phi$'s, and
finding the complete
set of these solutions is equivalent
to solve the so-called reduction equations \cite{zimmermann1},
$\beta_{g} \,(d g_{i}/d g) =\beta_{i}~,~i=1,\cdots,N$,
where $g$ and $\beta_{g}$ are the primary
coupling and its $\beta$-function.
Using all the $(N-1)\,\Phi$'s to impose RGI relations,
one can in principle
express all the couplings in terms of
a single coupling $g$.
 The complete reduction,
which formally preserve perturbative renormalizability,
 can be achieved by
demanding a 
 power series solution,
where the uniqueness of such a power series solution
can be investigated at the one-loop level \cite{zimmermann1}.
The completely reduced theory contains only one
independent coupling with the
corresponding $\beta$-function.
In supersymmetric Yang-Mills theories with
a simple gauge group, something more  drastic can happen;
the vanishing of the $\beta$-function
 to all orders in perturbation theory,
if all the one-loop anomalous dimensions of the matter fields
in the completely and uniquely reduced
theory vanish identically.

This possibility of coupling unification is
attractive, but  it can be too restrictive and hence
unrealistic. To overcome this problem, one  may use fewer $\Phi$'s
as RGI constraints.

\subsection { The $SU(5)$ Finite Unified Theory [3] } 

This is a $N=1$ supersymmetric Yang-Mills theory  based on
 $SU(5)$  which contains one ${\bf 24}$,
four pairs of (${\bf 5}+\overline{{\bf 5}}$)-Higgses and
 three
($\overline{{\bf 5}}+{\bf 10}$)'s
for three fermion generations. It has been done a complete reduction
 of the dimensionless parameters of the theory in favour of the gauge
 coupling $g$ and the unique power series 
solution \cite{mondragon2,kubo2}
 corresponds to 
the Yukawa matrices
without intergenerational mixing, and yields
in the one-loop approximation
$g_{t}^{2} =g_{c}^{2}=g_{u}^{2}=(8/5) g^2~,~
g_{b}^{2} =g_{s}^{2} =g_{d}^{2}=
g_{\tau}^{2} =g_{\mu}^{2} =g_{e}^{2}=(6/5) g^2$,
where $g_i$'s stand  for the Yukawa couplings.
\begin{table*}
\begin{center}\caption{Representative predictions of the FUT
 $SU(5)$}\label{table-fut}
\vspace{0.2cm}
\begin{tabular}{|c|c|c|c|c|c|}
\hline
$m_{\rm SUSY}$ [GeV]   &$\alpha_{3}(M_Z)$ &
$\tan \beta$  &  $M_{\rm GUT}$ [GeV]
 & $m_{b} $ [GeV]& $m_{t}$ [GeV]
 \\ \hline
$500$ & $0.118$  & $54.2$ & $1.45 \times 10^{16}$
 & $5.1$ & $184.4$ \\ \hline
\end{tabular}
\end{center}
\end{table*}
At first sight, this GYU seems to lead
to unacceptable predictions of the fermion masses.
But this is not the case, because each generation has
an own pair of ($\overline{{\bf 5}}+{\bf 5}$)-Higgses
so that one may assume that
after the diagonalization
of the Higgs fields  the effective theory is
exactly MSSM, where the pair of
its Higgs supermultiplets mainly stems from the
(${\bf 5}+\overline{{\bf 5}} $) which
couples to the third fermion generation.
(The Yukawa couplings of the first two generations
can be regarded as free parameters.)

\subsection{ The minimal supersymmetric $SU(5)$ model [4]}

The field content is minimal. Neglecting the
CKM mixing, one starts with six
Yukawa and two Higgs
couplings. We then require
GYU to occur among the Yukawa couplings of the third
generation and the gauge coupling.
We also require the theory to
be completely asymptotically free.
In the one-loop approximation, the GYU yields
$g_{t,b}^{2} ~=~\sum_{m,n=1}^{\infty}
\kappa^{(m,n)}_{t,b}~h^m\,f^n~g^2$ 
($h$ and $f$ are related
to the Higgs couplings). Where 
$h$ is allowed to vary from
$0$ to $15/7$, while $f$ may vary from $0$ to a maximum which depends
on $h$ and vanishes at $h=15/7$.
As a result, we obtain\cite{mondragon1,kubo2}:
$ 0.97\,g^2 \lsim g_{t}^{2} \lsim 1.37\,g^2~,~
~0.57\,g^2 \lsim g_{b}^{2}=g_{\tau}^{2}  \lsim 0.97\,g^2$.
We found \cite{kmoz2,kubo2} that consistency with proton decay requires
$g_t^2,~g_b^2$ to be very close to the left hand side values in the
inequalities. 

\vspace{0.2cm}
In all of the analyses above,
 we have  assumed that
it is possible to arrange
the susy mass parameters along with
the soft breaking terms in such a way that
the desired symmetry breaking pattern
 really occurs, all the superpartners are
unobservable at present energies,
and so forth.
To simplify our numerical analysis  we have also assumed a
unique  threshold $m_{\rm SUSY}$ for all
the  superpartners.
Using the updated experimental data on the SM parameters, we have
re-examined the $m_t$ prediction of the two GYU $SU(5)$ models
described above \cite{kmoz2,kubo2}.  They predict
$m_t = (183 + \delta ^{MSSM}_{m_t} \pm 5)$ GeV (FUT) and
$m_t = (181 + \delta ^{MSSM}_{m_t} \pm 3)$ GeV (Min. SUSY $SU(5)$),
where $\delta ^{MSSM}_{m_t}$ stands for the MSSM threshold
corrections of few \%.

\begin{table*}
\begin{center}\caption{Representative presdictions of the minimal SUSY
    $SU(5)$}\label{table-afut} 
\vspace{0.2 cm}
\begin{tabular}{|c|c|c|c|c|c|c|c|}
\hline
$m_{\rm SUSY}$ [GeV]
& $g_{t}^{2}/g^2$ & $g_{b}^{2}/g^2$&
$\alpha_{3}(M_Z)$ &
$\tan \beta$  &  $M_{\rm GUT}$ [GeV]
 & $m_{b} $ [GeV]& $m_{t}$ [GeV]
  \\ \hline
$500$ & $0.97$& $0.57$ & $0.118$  & $47.7$ & $1.39\times10^{16}$
  & $5.3$  & $178.9$  \\ \hline
\end{tabular}
\end{center}
\end{table*}

\section{Discussion and Conclusions}

The gauge-Yukawa unification we have been 
proposing is based on the RGI relations among gauge
and Yukawa couplings in GUTs beyond the unification scale.
They correspond to a
``technically natural fine tuning''
and could point to
a unified scheme at the Planck scale.
We have not assumed  the precise nature of the unified
scheme, but
 these relations could be obtained at the
Planck scale and being RGI survive down to the GUT scale. 
We have found  \cite{kubo2,mondragon2,mondragon1,kmoz2} that 
 two supersymmetric $SU(5)$ GUTs with GYU can
 predict the bottom and top
quark masses in accordance with the recent experimental data.
This means that the top-bottom hierarchy 
could be
explained in these models,
 in a similar way as 
the hierarchy of the gauge couplings of the SM
can be explained if one assumes  the existence of a unifying
gauge symmetry at $M_{\rm GUT}$. 
We would like to emphasize that the GYU scenario  is the most predictive
scheme as far as the mass of the top quark is concerned.
We also have found  \cite{kubo2,kmoz2}  that
the lowest value of the infrared quasi-fixed-point value
 of $m_t$  for large $\tan \beta$ is $\sim 188$ GeV.  Comparing this with
the experimental value $m_t = (175.6 \pm 5.5)$ GeV \cite{topmass} 
we may conclude
that the present data on $m_t$ cannot be explained from the infrared
quasi-fixed-point behaviour alone if 
$\tan \beta \gsim 2$.

Further, we have extended our approach \cite{kubo3} and have been searching 
for RGI relations among the soft supersymmetry-breaking (SSB)
parameters of the models. Preliminary results in this case of the model
discussed in $2.2$ show that the low energy SSB sector of the theory
contains a single arbitrary parameter, the unified gaugino mass $M_g$,
which characterizes the scale of the supersymmetry breaking.
In turn the lightest Higgs mass is predicted to be $\sim 120$ GeV,
while the lightest supersymmetric particle is found to be a
neutralino of $\sim 220$  GeV for $M_g (M_{GUT}) \sim 0.5$ TeV
\cite{kubo3}. Moreover, it has been recently found \cite{kawamura1} that
 RGI SSB scalar masses in gauge-Yukawa unified models 
satisfy a universal  sum
rule,  which coincides with those obtained
in a certain class of  superstring models in which
the origin of the sum rule has a symmetry interpretation,
target-space duality.

\end{document}